\documentclass[12pt]{article}

\usepackage{epsfig}
\usepackage{amsfonts}
\usepackage{amssymb,amsmath}

\newcommand{\mbf}[1]{{\boldsymbol {#1} }}
\newcommand{\complex}{{\mathbb C}} %% complex numbers
\newcommand{\zed}{{\mathbb Z}} %% integers
\newcommand{\nat}{{\mathbb N}} %% natural numbers
\newcommand{\real}{{\mathbb R}} %% real numbers
\newcommand{\rat}{{\mathbb Q}} %% rational numbers
\newcommand{\torus}{{\mathbb T}}
\def\e{{\,\rm e}\,}

\newcommand{\vq}{{\mbf q}}

\newcommand{\vnu}{{\mbf\nu}}

\def\ii{{\,{\rm i}\,}}
\def\dd{{\rm d}}
\def\Li{{\rm Li}}

\newcommand{\sect}[1]{\noindent{\bf {#1}.} }

\newcommand{\eq}{\begin{equation}}
\newcommand{\eqend}{\end{equation}}
\newcommand{\eqa}{\begin{eqnarray}}
\newcommand{\nonueqa}{\begin{eqnarray*}}
\newcommand{\eqaend}{\end{eqnarray}}
\newcommand{\nonueqaend}{\end{eqnarray*}}

\newcommand{\bma}[1]{\begin{array}{#1}}
\newcommand{\ema}{\end{array}}
\newcommand{\bc}{\begin{center}}
\newcommand{\ec}{\end{center}}

\def\appendix#1{\addtocounter{section}{1}\setcounter{equation}{0}
\renewcommand{\thesection}{\Alph{section}}
\section*{Appendix \thesection\protect\indent \parbox[t]{11.715cm} {#1}}
\addcontentsline{toc}{section}{Appendix \thesection\ \ \ #1} }

\newif\ifold             \oldtrue

\def\e{{\,\rm e}\,}

\def\beq{\begin{equation}}
\def\eeq{\end{equation}}
\def\bea{\begin{eqnarray}}
\def\eea{\end{eqnarray}}
\def\bd{\begin{displaymath}}
\def\ed{\end{displaymath}}

\newtheorem{theorem}{Theorem}
\newtheorem{proposition}{Proposition}

\begin{document}

\begin{flushright}
\baselineskip=12pt
HWM--04--16\\
EMPG--04--06\\
hep--th/0408055\\
\hfill{ }\\
August 2004
\end{flushright}

\begin{center}

\baselineskip=20pt

{\Large\bf Two-dimensional Yang-Mills theory\\[1mm] and moduli spaces of
  holomorphic differentials}

\baselineskip=12pt

\vspace{0.5 cm}

{\large L.~Griguolo$^a$, D.~Seminara$^b$ and R.J.~Szabo$^c$}\\

\vspace{0.4 cm}

{\it $^a$ Dipartimento di  Fisica, Universit\`a  di Parma,
INFN Gruppo Collegato di Parma\\
Parco Area delle Scienze 7/A, 43100 Parma, Italy\\
{\tt griguolo@fis.unipr.it}}\\[3mm]

{\it $^b$ Dipartimento di Fisica, Polo Scientifico Universit\`a di
  Firenze\\ INFN Sezione di Firenze\\
Via  G. Sansone 1, 50019 Sesto Fiorentino, Italy\\
{\tt seminara@fi.infn.it}}\\[3mm]

{\it $^c$ Department of Mathematics, Heriot-Watt University\\ Scott
  Russell Building, Riccarton,
Edinburgh EH14 4AS, U.K.\\{\tt R.J.Szabo@ma.hw.ac.uk}}

\end{center}

\begin{abstract}

\baselineskip=12pt

\noindent
We describe and solve a double scaling limit of large $N$ Yang-Mills theory
on a two-dimensional torus. We find the exact strong-coupling
expansion in this limit and describe its relation to the conventional
Gross-Taylor series. The limit retains only the chiral sector of the
full gauge theory and the coefficients of the expansion determine the
asymptotic Hurwitz numbers, in the limit of infinite winding number,
for simple branched coverings of a torus. These numbers are computed
exactly from the gauge theory vacuum amplitude and shown to
coincide with the volumes of the principal moduli spaces of
holomorphic differentials. The string theory interpretation of the
double scaling limit is also described.

\end{abstract}

\baselineskip=14pt

\sect{1. Introduction} Two-dimensional Yang-Mills theory is an exactly
solvable quantum field theory which has over the years found a
multitude of both physical and mathematical applications. On the
physical side, it is the first example of a gauge theory which can be
reformulated precisely as a string theory~\cite{gross1}. It is also
deeply connected to integrable systems~\cite{gorsky1} and conformal
field theory~\cite{douglas}, and has recently been shown to be
equivalent to topological string theory on certain non-compact
Calabi-Yau threefolds~\cite{vafa1} and to reproduce the counting of
instantons in four-dimensional gauge theories with $\mathcal{N}=2$
supersymmetry~\cite{MMO1}. On the mathematical side, the partition
function is the generating function for intersection indices on the
symplectic moduli spaces of flat connections on Riemann
surfaces~\cite{witten1} and for the orbifold Euler characters of
Hurwitz moduli spaces of branched covers with deep links to
topological field theory in two dimensions~\cite{CMR1}.

In this paper we will describe some new applications of Yang-Mills
theory on a two-dimensional torus which is based on a double
scaling limit of the quantum gauge theory at large $N$. Starting from the exact
semi-classical expansion of the partition function, we derive the weak-coupling
expansion dual to the Gross-Taylor series and compute explicitly the
series originally obtained in~\cite{douglas,Rudd1} using the conformal
field theory approach. Our approach is based on the exact solutions of
certain saddle-point equations in the zero-instanton sector of the
two-dimensional gauge theory. We then define a new limit which
captures only one chiral sector of the gauge theory and gives a
 geometrical meaning to the expansions unveiled
in~\cite{Rudd1}. In this limit, only string states of infinite winding
number contribute to the vacuum amplitude which thereby becomes the
generating function for the asymptotics of Hurwitz numbers of simple
branched coverings of a torus. Our saddle-point technique provides a
very quick and efficient way to extract exact formulas for these
asymptotics which avoids the cumbersome combinatorial techniques
usually employed in the mathematics literature. With this realization,
we then go on to show that the partition function in the limit
actually serves as a generating function for the volumes of moduli
spaces of holomorphic differentials on Riemann surfaces. The beauty of
this realization is that the observables of two-dimensional Yang-Mills
theory, such as Wilson loops and Polyakov lines, could provide
generating functions for intersection indices, or other geometrical
quantities, on these moduli spaces. We will also briefly describe the
string theoretic meaning of the limit and argue that it is more akin
to a string representation of noncommutative gauge theory in two
dimensions. Further details and applications will be given
in~\cite{inprep}.

\bigskip

\sect{2. Chiral Gross-Taylor series on the torus} The chiral partition
function of $U(N)$ Yang-Mills theory on a rectangular torus $\torus^2$
of area $A$ is a variant of the usual Migdal strong-coupling expansion
given by~\cite{gross1}
\beq
Z^+(\lambda,N)=\sum_{R\in{\rm Rep}^+(U(N))}\e^{-\frac{\lambda
}{N}\,C_2(R)} \ ,
\label{chiralsumR}\eeq
where $\lambda=g^2A\,N/2$ is the dimensionless 't~Hooft coupling
constant and $C_2(R)$ is the quadratic Casimir eigenvalue in the
irreducible representation $R=R(Y)$ of the gauge group $U(N)$. The
restriction to chiral representations means considering Young tableaux
$Y$ with positive number of boxes and dropping the constraint that the
number of rows be less than the rank $N$. In so doing, one essentially assumes
that only representations with small numbers of boxes (compared to
$N$) are relevant in the large $N$ limit, the others being
exponentially suppressed in $N$~\cite{gross1}. More generally, a class
of representations with box numbers of order $N$, whose contribution
is not exponentially damped, can be found and used to construct the
anti-chiral sector. A non-chiral coupled expansion,
incorporating both contributions, has been presented in~\cite{gross1}
for the case of an $SU(N)$ gauge group and is widely accepted as the
complete large $N$ description of the gauge theory.

By using the explicit form of the quadratic Casimir, the
sum over Young diagrams may be carried out, and the free energy
$F^+(\lambda,N)=\ln Z^+(\lambda,N)$ can be written as the asymptotic
$\frac1N$ expansion~\cite{gross1}
\beq
F^+(\lambda,N)=\sum_{h=1}^\infty\frac1{N^{2h-2}}\,
F^+_h(\lambda) \ , ~~ F^+_h(\lambda)=\lambda^{2h-2}\,
\sum_{n=1}^\infty\omega_h^{n}~\e^{-n\,\lambda} \ .
\label{FglambdaA}\eeq
The non-negative integers $\omega_h^n$ are called simple Hurwitz
numbers and they give the number of (topological classes of)
$n$-sheeted holomorphic covering maps without folds to the torus
$\torus^2$. They count the number of maps from a closed, connected and oriented
Riemann surface of genus $h$ to the torus with winding number $n$ and $2h-2$
simple branch points. This expansion contains a
Nambu-Goto factor $\e^{-n\,\lambda}$ as well as a volume factor
$\lambda^{2h-2}$ from the moduli space integration over the positions
of the branch points. In other words,
the strong coupling expansion for the chiral free energy of $U(N)$ gauge
theory on $\torus^2$ in the 't~Hooft limit is the generating function
for the simple Hurwitz numbers, and the expansion (\ref{FglambdaA}) can be
identified as the partition function of a closed string theory with torus
target space. The coupling and tension are given by $g_s=\frac1N$ and
$T=\frac1{2\pi\,\alpha'}=\frac\lambda A$.

One of the most interesting properties of the QCD$_2$ string partition
function is that the contributions (\ref{FglambdaA}) are quasi-modular
forms on the elliptic curve whose K\"ahler class is dual to the modulus
$\tau=-\frac{\lambda}{2\pi\ii}$, i.e. they are
polynomials in the basic holomorphic Eisenstein series $E_2(\tau)$,
$E_4(\tau)$ and $E_6(\tau)$, with $F^+_h(\tau)$, $h\geq2$ of
weight $6h-6$ under the full modular group
$PSL(2,\zed)$~\cite{Rudd1,dijkgraaf}. This was first observed
in~\cite{Rudd1} by direct inspection of the Feynman diagram expansion
for the free energy within the conformal field theory approach to large
$N$ Yang-Mills theory proposed in~\cite{douglas}. A rigorous proof was
subsequently presented in~\cite{dijkgraaf} directly from the
equivalent free fermion representation of the partition function.
The quasi-modular character of the $F^+_h(\tau)$
and their computability through the Feynman diagram expansion in a
Kodaira-Spencer field theory confirm two general predictions of the
mirror symmetry program in the special case of elliptic
curves~\cite{bcov}.

Explicit formulas for the free energy contributions up to genus $h=8$
are given in~\cite{Rudd1} (see also~\cite{MSS1}). In a suitable basis
for the modular forms, they generally assume the forms
\bea
F_1^+(\lambda)&=&-\epsilon^{~}_{\rm F}-\ln\eta(\tau) \ , \nonumber
\\F_h^+(\lambda)&=&\frac{\lambda^{2h-2}}{(2h-2)!\,\rho_h}\,
{}~\sum_{k=0}^{3h-3}~\sum_{\stackrel{\scriptstyle l,m\geq0}
{\scriptstyle 2l+3m=3h-3-k}}s_{kl}~E_2^{~}(\tau)^k\,E_2^\prime(\tau)^l
\,E_2^{\prime\prime}(\tau)^m
\label{FgUNexpl}\eea
with $h\geq2$ and $\rho_h,s_{kl}\in\nat$, where $\epsilon^{~}_{\rm
  F}=-\frac\lambda{12}\,N^2(N^2-1)$ is the Fermi energy and
  $\eta(\tau)$ is the Dedekind function. By using the modular
  transformation properties of the quasi-modular forms appearing in
  (\ref{FgUNexpl}), it is possible to cast the string representation
(\ref{FglambdaA}) in the weak-coupling regime $\lambda\to0$ of the
gauge theory and analyse the {}{Douglas-Kazakov type}
singularity at zero area whereby a condensation of instantons in
the vacuum occurs~\cite{dk1}. It was argued in~\cite{Rudd1}
that the contributions are of the  form
\beq
F_h^+(\lambda)=\lambda^{2h-2}\,
\sum_{k=3h-3}^{4h-3}\frac{r_{k,h}~\pi^{2(k-3h+3)}}{\lambda^k}
+O\left(\e^{-1/\lambda}\right) \label{Fggeneric}
\eeq
where $r_{k,h}\in\rat$ are related to the simple Hurwitz numbers.
{}{The structure of the small area singularities is very
peculiar and it is not the most general one expected from the
quasi-modular behaviour.} In
this paper we will elucidate the geometrical meaning of
the rational numbers $r_{k,h}$. Explicit expressions up to genus
$h=6$, reflecting this structure, are found in~\cite{Rudd1}.

\bigskip

\sect{3. Saddle-point solution} The proper setting for the study of
weak-coupling expansions of the sort (\ref{Fggeneric}) is the
instanton expansion, which for $U(N)$ Yang-Mills theory on $\torus^2$
is given by~\cite{grig}
\bea
Z(\lambda,N)&=&(-1)^N~\e^{-\epsilon^{~}_{\rm F}}\,\sum_{
\vnu\in\nat_0^{N}\,:\,\sum_kk\,\nu_k=N}~
\prod_{k=1}^{N}\,\frac{(-1)^{\nu_k}}{\nu_k!}\,\left(
\frac{\pi^2N}{k^3\lambda}\right)^{\nu_k/2}\nonumber\\&&\times\,
\sum_{\vq\in\zed^{|\vnu|}}(-1)^{(N-1)\sum_kq_k}\,\exp\left[-
\frac{\pi^2N}{\lambda}\,\sum_{l=1}^N\frac1l\,\sum_{j=1+\nu_1+\dots+
\nu_{l-1}}^{\nu_1+\dots+\nu_l}q_j^2\right] \ ,
\label{ZUNinstexp}\eea
where we have defined $\nu_0:=0$ and the total number of partition components
$|\vnu|:=\nu_1+\dots+\nu_{N}$. In (\ref{ZUNinstexp}) it is
understood that if some $\nu_k=0$ then
$q_{1+\nu_1+\dots+\nu_{k-1}}=\dots=q_{\nu_1+\dots+\nu_k}=0$. This
result is derived by an explicit Poisson resummation of the full
(non-chiral) Migdal expansion of the gauge theory. It represents the
exact semi-classical expansion of the gauge theory path integral,
written as a sum over unstable instantons.

By defining the sequence of functions
\beq
\Xi_k(\lambda):=\sum_{q=-\infty}^\infty(-1)^{(N-1)q}~
\e^{-\frac{\pi^2N}\lambda\,\frac{q^2}k}
\label{calFkdef}\eeq
and resolving the constraint on the sum over
partitions $\vnu$ in (\ref{ZUNinstexp}) explicitly by using a contour
integral representation, we arrive at the formula
\beq
Z(\lambda,N)=\e^{-\epsilon^{~}_{\rm F}}\,
\oint\frac{\dd z}{2\pi\ii z^{N+1}}~
\exp\left[-\sqrt{\frac{\pi^2N}\lambda}~\sum_{k=1}^{\infty}\,\Xi_k
(\lambda)~\frac{(-z)^k}{k^{3/2}}\right]
\label{ZUNcontint}\eeq
where the contour of integration encircles the origin $z=0$ of the
complex $z$-plane and lies in the open unit disc $|z|<1$ in order to ensure
convergence of the integrand. This representation of the weak-coupling
partition function will be the key to extracting its form in the
desired large $N$ scaling limits. It can also be used to derive a
concise resummation of the Migdal expansion. Applying the Poisson
resummation formula to the sequence of functions (\ref{calFkdef}), we
may sum the series over $k$ explicitly and bring (\ref{ZUNcontint})
into the equivalent strong-coupling form
\beq
Z(\lambda,N)=\e^{-\epsilon^{~}_{\rm F}}\,
\oint\frac{\dd z}{2\pi\ii z^{N+1}}~
\prod_{n=-\infty}^{\infty}\,\left(1+z~\e^{-\frac\lambda N\,
(n-n^{~}_{\rm F})^2}\right) \ ,
\label{ZUNstrongcoupl}\eeq
where $n_{\rm F}^{~}=\frac{N-1}2$ is the Fermi level.
The contour integration now implements the constraint
that the number of rows in the Young diagrams be bounded from above by
the rank $N$ of the gauge group. {}{Eq. (\ref{ZUNstrongcoupl}) agrees
  with the representation derived in \cite{Panzeri} by direct group
  theory arguments and it} is also similar to the contour integral
formula obtained in~\cite{douglas} from the free fermion
representation of the partition function.

We will now compute the partition function
(\ref{ZUNcontint}) in the large $N$ limit with the 't~Hooft coupling constant
$\lambda=g^2A\,N/2$ held fixed, and compare it to the string
representation of the previous section. In the weak coupling limit,
the higher instanton contributions to the function (\ref{calFkdef})
are of order $\e^{-N/\lambda}$ and hence can be neglected to a
first-order approximation at $N\to\infty$. We may thereby focus on the
zero-instanton sector with vanishing magnetic charge $q=0$, and write
the vacuum amplitude as
\beq
{\cal Z}(\lambda,N)=\e^{-\epsilon^{~}_{\rm F}}\,
\oint\frac{\dd z}{2\pi\ii z}~\exp\left[\mbox{$-N\ln z-\sqrt{\frac{\pi\,N}
{\lambda}}~\Li_{3/2}(-z)$}\right] \ ,
\label{largeNstringZUN}\eeq
where generally $\Li_\alpha(z):=\sum_{k\in\nat}z^k/k^\alpha$
denotes the polylogarithm function of index $\alpha$. We will compute
the integral (\ref{largeNstringZUN}) in the large $N$ limit by means
of saddle-point techniques. The saddle-point equation is
\beq
\Li_{1/2}(-z)=\mbox{$-\sqrt{\frac{N\,\lambda}{\pi}}$} \ .
\label{saddlepteqGT}\eeq
The function $\Li_{1/2}(-z)$ is a slowly
decreasing negative function for $z\geq1$, behaving as
$\Li_{1/2}(-z)\simeq-2\,\sqrt{\ln(z)/\pi}$ for $z$ real with
$z\to\infty$. Thus a solution to (\ref{saddlepteqGT})
always exists and is located at large $z\in[-1,\infty)$.

The saddle-point equation (\ref{saddlepteqGT}) can be solved to any
order in $N$ by using the large $x$ asymptotic expansion of the
polylogarithm function~\cite{inprep}
\beq
\Li_\alpha\left(-\e^x\right)=2\,\sum_{k=0}^\infty
\frac{\left(1-2^{2k-1}\right)\,B_{2k}~\pi^{2k}}
{(2k)!~\Gamma(\alpha+1-2k)}~x^{\alpha-2k} \ ,
\label{Lialphaasympt}\eeq
where $B_{2k}\in\rat$ are the Bernoulli numbers. Corrections to this
formula are of order $\e^{-x}$. We seek a solution of the form
$z=z_*=\e^{x_*}$ with $x_*$ admitting an asymptotic $\frac1N$
expansion
\beq
x_*=x_{-1}\,N+\sum_{k=0}^\infty\frac{x_k}{N^k} \ .
\label{xstaransatz}\eeq
One can now proceed to obtain the coefficients
$x_k$ of the saddle-point solution recursively in powers of $\frac1N$
by substituting (\ref{xstaransatz}) and (\ref{Lialphaasympt}) into
(\ref{saddlepteqGT}), and the result of a straightforward iterative
evaluation up to order $1/N^{11}$ reads
\beq
x_*=\mbox{$\frac{N\,\lambda}4+\frac{\pi^2}{3N\,\lambda
}+\frac{16\pi^4}{9\left(N\,\lambda\right)^3}
+\frac{448\pi^{6}}{9\left(N\,\lambda\right)^5}
+\frac{1254656\pi^{8}}{405\left(N\,\lambda\right)^7}+
\frac{406598656\pi^{10}}{1215\left(N\,\lambda\right)^9}+
\frac{67556569088\pi^{12}}{1215\left(N\,\lambda\right)^{11}}+
O\left(\frac{1}{N^{13}}\right)$} \ .
\label{xstarGTN11}\eeq
The systematic vanishing of the $x_k$ for even powers of $\frac1N$ is
exactly what is expected from the form of the Gross-Taylor expansion.

Armed with the solution (\ref{xstarGTN11}) of the saddle-point
equation, we can now proceed to compute the free energy ${\cal
  F}(\lambda,N)=\ln{\cal Z}(\lambda,N)$ by
parametrizing the integration variable in (\ref{largeNstringZUN}) as
$z=\e^{x_*+x}$, where the variable $x$ contains contributions from
fluctuations about the saddle-point value which can be integrated over
all $x\in\real$ since the deviations from the results computed with the correct
integration domain appropriate to (\ref{largeNstringZUN}) will be
exponentially suppressed {}{in the small area limit}. By using
(\ref{Lialphaasympt}), a numerical evaluation up to order $1/N^{10}$
using {\sl Mathematica} yields
\bea
{\cal F}(\lambda,N)&=&\mbox{$\frac{\ln(-2\pi\,\lambda)}2
+\frac{\pi^2}{6\lambda}+\left(\frac{2}{3\lambda
}-\frac{2\pi^2}{3\lambda^2}+\frac{8
\pi^4}{45\lambda^3}\right)\,\frac1{N^2}
+\left(-\frac{8}{\lambda^2}+\frac{16\pi^2}
{\lambda^3}-\frac{100\pi^4}{9
\lambda^4}+\frac{224\pi^6}{81
\lambda^5}\right)\,\frac1{N^4}$}\nonumber\\&&\mbox{$
+\,\left(\frac{2272}{9\lambda^3}-\frac{2272\pi^2}
{3\lambda^4}+\frac{8096\pi^4}{9
\lambda^5}-\frac{41504
\pi^6}{81\lambda^6}+\frac{48256\pi^8}{405
\lambda^7}\right)\,\frac1{N^6}$}\nonumber\\&&\mbox{$
+\,\left(-\frac{13504}{\lambda^4}+\frac{54016
\pi^2}{\lambda^5}-\frac{834304\pi^4}
{9\lambda^6}+
\frac{7010816\pi^6}{81\lambda^7}-\frac{17887904
\pi^8}{405\lambda^8}+\frac{11958784\pi^{10}}
{1215\lambda^9}\right)\,\frac1{N^8}$}\nonumber\\&&\mbox{$
+\,\left(\frac{15465472}{15\lambda^5}-
\frac{15465472\pi^2}{3\lambda^6}+
\frac{105156608\pi^4}{9
\lambda^7}-\frac{418657280\pi^6}
{27\lambda^8}\right.$}\nonumber\\&&\mbox{$+\left.
\frac{572409344\pi^8}{45\lambda^9}-
\frac{2467804672\pi^{10}}{405
\lambda^{10}}+\frac{33778284544\pi^{12}}{25515
\lambda^{11}}\right)\,\frac1{N^{10}}+
O\left(\frac1{N^{12}}\right)$} \ .
\label{largeNfreenUNexp}\eea
The expression (\ref{largeNfreenUNexp}) matches {\it precisely}
eqs.~(3.43)--(3.47) in~\cite{Rudd1} which were obtained using the
conformal field theory representation of large $N$  two-dimensional
Yang-Mills theory. The saddle-point evaluation of the zero-instanton sector
thereby reproduces all non-exponentially suppressed terms in the weak-coupling
expansion of the chiral $U(N)$ free energy, giving the correct
rational numbers $r_{k,h}$ appearing in (\ref{Fggeneric}). An
important ingredient in this reproduction is the cancellation of
the ground state energy $\epsilon^{~}_{\rm F}$, since this term would
otherwise dominate the series in the large $N$ limit. In particular,
the expansion starts at order $N^0$ and there is no spherical
contribution, consistent with the fact that there are no unfolded
coverings of a torus by a sphere. The expansion
(\ref{largeNfreenUNexp}) also contains the correct leading modular
dependence of the Dedekind function coming from the genus~$1$ free
energy. Let us also point out three particularly noteworthy features
of our derivation of the formula (\ref{largeNfreenUNexp}). First of
all, it is a highly accurate check of the expansion obtained
in~\cite{Rudd1} by a completely independent method. Secondly, it does not rely
on any group theory or conformal field theory techniques and is
obtained directly in the instanton representation. As far as we are
aware, this is the first time that stringy quantities are computed in
two dimensional Yang-Mills theory directly from the weak-coupling expansion,
which is the most natural one from a field theoretic
point of view. This is possible due to the absence of a large $N$
phase transition at finite area on the torus, which would otherwise
prohibit the recovery of stringy features from weak-coupling
data~\cite{dk1}. Finally, we note that the dynamics of the
zero-instanton sector is surprisingly rich, encoding properly the
anticipated string characteristics. This is related
to the underlying structure of the topological string theory governing
the weak-coupling limit~\cite{CMR1,bcov}.

However, despite this remarkable agreement, the computation above
reproduces only the {\it chiral} part of the full $U(N)$ gauge
theory. The saddle-point technique has not picked up the coupling to
the anti-chiral contributions, representing the anti-holomorphic
sector of the closed string theory. Part of
the problem can be traced back to the fact that the argument of the exponential
integrand in (\ref{largeNstringZUN}) does not admit a nice large $N$
scaling. Moreover, we eventually have to face
up to the problem of evaluating the non-zero higher instanton contributions. It
is very likely that, in spite of their exponential suppression order
by order in $\frac1N$, their collective behaviour will be crucial to
recover the complete string expansion. The problem of how much information
should be carried by the higher instantons in order to reproduce the
full string partition function will be addressed elsewhere~\cite{inprep}.

\bigskip

\sect{4. Double scaling limit} Motivated by the analysis of the
previous section, we will now analyse the $U(N)$ gauge theory in a
large $N$ limit wherein the saddle-point technique can capture the
entire relevant story. We will take the
limit $N\to\infty$ while keeping fixed the new scaled coupling constant
$\mu:=\mbox{$\frac{1}{\pi}$}\,N\,\lambda=\mbox{$\frac{1}{2\pi}$}\,N^2g^2A$.
The partition function then assumes the form
\beq
\hat{\cal Z}(\mu,N)=\e^{\frac{\pi\,N\,\mu}{12}}\,
\oint\frac{\dd z}{2\pi\ii z}~
\exp\left[-N\left(\mbox{$\ln z+\frac1{\sqrt\mu}~\Li_{3/2}(-z)$}
\right)\right]:=\oint\frac{\dd z}{2\pi\ii z}~\e^{N\,\hat F(z,\mu)}
\label{partfndoubleUN}\eeq
and hence has a nice large $N$ limit. Corrections to this expression
from higher instanton configurations are of order $\e^{-N^2/\mu}$ and
could be completely suppressed in the $\frac1N$ expansion. In fact, at
$N=\infty$ the vacuum amplitude is given by the leading planar term $\hat{\cal
  Z}(\mu,N)=\e^{N\,\hat F(z_*,\mu)}$ in the $\frac1N$ expansion of
the integral (\ref{partfndoubleUN}), which can be rigorously computed
in the saddle-point approximation. We refer to this new limit of
the gauge theory as the ``double scaling limit''. It is very different
from the conventional planar large $N$ limit used to derive the
Gross-Taylor expansion.

Starting from (\ref{partfndoubleUN}), we derive the saddle-point
equation \beq \Li_{1/2}(-z)=-\sqrt\mu \ ,
\label{saddleptdoubleUN}\eeq and we will show that it can be
solved exactly. In contrast to the conventional 't~Hooft limit,
this equation does not depend on $N$ and therefore its solution
$z=z_*(\mu)$ does not rely on any approximation in general. It can
be simply written as $z_*(\mu)=-\Li^{-1}_{1/2}(-\sqrt\mu\,)$, the
inverse function being uniquely defined in the region of interest
thanks to the monotonic behaviour of the polylogarithm function
$\Li_{1/2}(-z)$. The double scaling free energy can then be
evaluated as $\hat{\cal F}(\mu,N)=N\,\hat F(z_*(\mu),\mu)$. We can
write down an {\it exact} relation that gives it directly as a
primitive of the position of the saddle point. By parametrizing
the saddle point solution as before in the form
$z_*(\mu)=\e^{x_*(\mu)}$ and defining $y:=\sqrt{\mu}$, one can
easily derive the equation \beq \hat
F\bigl(\e^{x_*(\mu)}\,,\,\mu\bigr) =\frac{1}{\sqrt{\mu}}~
\int_{\sqrt{\mu}}^\infty\dd y~\left[x_*\left(y^2\right)-
\mbox{$\frac{\pi}{4}$}\,y^2\,\right]\ . \label{bella}\eeq This explicit
representation is useful because it exhibits the structure of the
free energy straightforwardly in terms of the properties of the
saddle point solution. It is a smooth function of $\mu>0$, with a
logarithmic singularity in the weak-coupling limit $\mu\to 0$
where the Douglas-Kazakov type phase transition takes place.

The physical meaning of the double scaling limit can be understood by
expanding the free energy for large $\mu$. It is in this regime that
we expect to see a relation with the string picture derived previously
in the 't~Hooft scaling limit. The strong-coupling saddle point
solution is given by the expansion (\ref{xstarGTN11}) which is
naturally written as a series in the double scaling parameter given by
\beq
x_*(\mu)=\pi\,\sum_{k=0}^{\infty}\,\frac{\xi_{2k-1}}{\mu^{2k-1}} \ .
\label{expa}\eeq
The double scaling free energy may then be computed directly from
(\ref{bella}) to get
\beq
\hat{\cal F}(\mu,N)=\pi\,N\,\left[\,\sum_{k=1}^{\infty}\frac{\xi_{2k-1}}
{4k-3}~\frac1{\mu^{2k-1}}+O\left(\e^{-\mu}\right)\right] \ ,
\label{freenDSdirect}\eeq
where we note the cancellation of the vacuum energy contribution. From
the explicit expression in (\ref{xstarGTN11}) the first few terms are
found to be given by
\beq
\hat{\cal F}(\mu,N)=2\pi\,N\,
\left[\mbox{$\frac{1}{6\mu}+\frac{8}{45\mu^3}+\frac{224}{81\mu^5}
+\frac{48256}{405\mu^7}+\frac{11958784}
{1215\mu^9}+\frac{33778284544}{25515
\mu^{11}}+O\left(\frac1{\mu^{13}}\right)+O\left(\e^{-\mu}\right)$}\right] \ .
\label{largeNfreenUNdouble}\eeq
Comparing with (\ref{largeNfreenUNexp}), we see that at
strong-coupling the double scaling limit has extracted the most
singular terms, as $\lambda\to0$, at each order of the original $\frac1N$
expansion. In other words, the double scaled gauge theory at
strong-coupling presents a resummation of the most singular terms in
the weak-coupling limit of the chiral Gross-Taylor string
expansion. The leading contribution to (\ref{Fggeneric}) in the double
scaling limit at $N=\infty$ is given at $k=4h-3$, and thus the general
form of this expansion can be written as in (\ref{freenDSdirect}) with
the rational numbers $r_{4h-3,h}$ completely determined by the
strong-coupling solution (\ref{expa}) of the saddle point equation as
$r_{4h-3,h}=\xi_{2h-1}/(4h-3)$.

Let us now elucidate the geometrical meaning of the rational numbers
$r_{4h-3,h}$. The $\lambda\to0$ behaviour of the series (\ref{FglambdaA}) is
controlled by the large $n$ asymptotics of the simple Hurwitz numbers
$\omega_h^n$. At fixed genus $h$, the singularities of the
free energy (\ref{FglambdaA}) as $\lambda\to
0$ are related to a power-like growth
$\omega_h^n\simeq\beta_h\,n^{\alpha_h}$ of the number of holomorphic
branched covering maps of $\torus^2$
with large winding number $n$, where $\alpha_h>0$ and
\beq
\beta_h=(\alpha_h+1)\,\lim_{N\to\infty}\,\frac1{N^{\alpha_h+1}}\,
\sum_{n=1}^N\omega_h^n \ .
\label{betahasympts}\eeq
The leading singularity of the series (\ref{FglambdaA}) as $\lambda\to
0$ is extracted by substituting these asymptotics to get
\beq
\lim_{\lambda\to0}\,F^+(\lambda,N)=
\sum_{h=1}^\infty\left(\frac{\lambda}N\right)^{2h-2}\,\beta_h~
\Li_{-\alpha_h}\bigl(\e^{-\lambda}\bigr) \ .
\label{FUNsingsubs}\eeq
In the limit $\lambda\to0$, we can substitute the leading singular
behaviour of the polylogarithm function
$\Li_{-\alpha_h}(z)\simeq\Gamma(\alpha_h+1)\,(-\ln z)^{-\alpha_h-1}$ for
$z\to1^-$~\cite{flajolet1}. Matching this to the expansion
(\ref{freenDSdirect}) of the free energy with $\lambda=\pi\,\mu/N$, we
find the power of the growth as the natural number $\alpha_h=4h-4$,
while the positive numbers $\beta_h$ are given by
$\beta_h=\xi_{2h-1}\,\pi^{2h}/(4h-3)!$. That the explicit
knowledge of the small area behaviour of the gauge theory allows one
to reconstruct the asymptotic forms of the simple Hurwitz numbers is
our first main result.
\begin{proposition}
The asymptotic expansion as $\mu\to\infty$ for the free energy
(\ref{freenDSdirect}) of $U(N)$ gauge theory on $\torus^2$ in the
double-scaling limit is the generating function for the asymptotic
Hurwitz numbers with
$$
\lim_{n\to\infty}\,
\omega_h^n=\frac{\pi^{2h}}{(4h-3)!}~\xi_{2h-1}~n^{4h-4} \ .
$$
\label{Hurwitzprop1}\end{proposition}

Thus the saddle-point equation for the zero-instanton double
scaling free energy solves the combinatorial problem of determining
the asymptotics of Hurwitz numbers. We will now compute the explicit
forms of the coefficients of the saddle-point expansion
(\ref{expa},\ref{freenDSdirect}) as polynomials in Bernoulli numbers,
and thereby write down the {\it exact} solution of the double scaling
gauge theory in the strong-coupling limit. For this, we set
$x_*=1/w^2$ and use (\ref{Lialphaasympt}) to write the saddle-point
equation (\ref{saddleptdoubleUN}) as
\beq
w=-\frac2{\sqrt\mu}\,\sum_{k=0}^\infty\frac{\left(1-2^{2k-1}\right)\,
B_{2k}~\pi^{2k}}{(2k)!~\Gamma\left(\frac32-2k\right)}~w^{4k}:=
\frac1{\sqrt\mu}~L(w) \ .
\label{saddleptTaylor}\eeq
The solution $w(\mu)$ of (\ref{saddleptTaylor}) can be found by means
of the Lagrange inversion formula. For our purposes it will be more
useful to employ the Burmann generalization, which may be represented
succinctly as the contour integral
\beq
G\bigl(w(\mu)\bigr)=\oint\frac{\dd z}{2\pi\ii}~G(z)~
\frac{1-\frac1{\sqrt\mu}\,L'(z)}{z-\frac1{\sqrt\mu}\,L(z)}
\label{Burmanngen}\eeq
where $G(z)$ is any analytic function. When $G(z)=z$, the formal
Taylor series expansion of the integrand of (\ref{Burmanngen}) in
powers of $\frac1{\sqrt\mu}$ reproduces the usual Lagrange inversion
formula. In our case, we should take
$G(z)=1/z^2$, but this cannot be directly inserted into the formula
(\ref{Burmanngen}) as it would introduce a spurious contribution from
the double pole at $z=0$. The simplest way to deal with this problem
is to subtract the undesired contribution by hand, and thereby write the
solution $x_*(\mu)$ of the saddle-point equation as
\beq
x_*(\mu)=\oint\frac{\dd z}{2\pi\ii z^2}~
\frac{1-\frac1{\sqrt\mu}\,L'(z)}{z-\frac1{\sqrt\mu}\,L(z)}-
\frac{L(0)\,L''(0)-\bigl(\sqrt\mu-L'(0)\bigr)^2}{L(0)^2} \ .
\label{xmucontint}\eeq

By formally expanding (\ref{xmucontint}) in powers of
$\frac1{\sqrt\mu}$, we arrive at the strong-coupling solution \beq
x_*(\mu)=-
\frac{L(0)\,L''(0)-\bigl(\sqrt\mu-L'(0)\bigr)^2}{L(0)^2}
-2\,\sum_{k=1}^\infty\frac1{(k+2)!\,k}\,
\left.\frac{\dd^{k+2}}{\dd z^{k+2}}L(z)^{k}\right|_{z=0}~
\frac1{\mu^{k/2}} \ . \label{xmuexpansion}\eeq {}From the
definition of the function $L(z)$ in (\ref{saddleptTaylor}), one
finds that the coefficients of the series in (\ref{xmuexpansion})
are non-zero only when $k=4h-2$ for some $h\in\nat$. Written in
the form (\ref{expa}), one computes $\xi_{-1}=\frac14$ and \beq
\xi_{2h-1}=-\frac{\pi^{2h-1}}{2h-1}~\sum_{\mbf
h\in\nat_0^{4h-2}\,:\,
\sum_kh_k=h}~\prod_{k=1}^{4h-2}\,\frac{\left(2-2^{2h_k}\right)\,
B_{2h_k}}{(2h_k)!~\Gamma\left(\frac32-2h_k\right)}
\label{xiexact}\eeq for $h\geq1$, and we have thereby found the
complete strong-coupling expansion of the double scaled gauge
theory. We can simplify the sum over ordered partitions $\mbf
h\in\nat_0^{4h-2}$ of the integer $h$ by reducing it to a sum over
partitions of $h$ into $m$ positive integers. By inserting~$0$
into all possible positions we obtain ${4h-2}\choose m$ partitions
of the original type in (\ref{xiexact}), and we find \bea
\xi_{2h-1}&=&-\frac{\pi^{2h-1}}{2h-1}\,\sum_{m=1}^h{4h-2\choose m}
\left(\frac{B_0}{\Gamma\left(\frac32\right)}\right)^{4h-2-m}\,
\sum_{\stackrel{\scriptstyle\mbf h\in\nat^m}
{\scriptstyle\sum_kh_k=h}}~\prod_{k=1}^m\,\frac{\left(2-2^{2h_k}\right)\,
B_{2h_k}}{(2h_k)!~\Gamma\left(\frac32-2h_k\right)}
\nonumber\\&=&-\sum_{m=1}^h\frac{(-1)^m\,2^{2h+m-1}\,(4h-3)!}
{(4h-2-m)!\,m!}~\sum_{\stackrel{\scriptstyle\mbf h\in\nat^m}
{\scriptstyle\sum_kh_k=h}}~\prod_{k=1}^m\,\frac{\left(2^{2h_k-1}-1
\right)\,(4h_k-3)!!~B_{2h_k}}{(2h_k)!} \ . \nonumber\\&&
\label{xireduce}\eea

Finally, by exploiting the symmetry of the second summand in
(\ref{xireduce}) we can reduce the sum over ordered partitions of $h$
with $m$ components to a sum over conjugacy classes and cycles of the
symmetric group $S_h$, i.e. over unordered partitions of $h$. An
unordered partition of $h$ is specified by $h$ non-negative integers
$\nu_k$ with $\sum_kk\,\nu_k=h$, while the condition that the
partition contain only $m$ parts is implemented by requiring that
$\sum_k\nu_k=m$. By inserting the combinatorial factor
$\frac{m!}{\nu_1!\cdots\nu_h!}$ which counts the number of different
ordered partitions that originate from the same unordered partition,
we may bring (\ref{xireduce}) into our final equivalent form.
\begin{theorem}
The coefficients of the strong-coupling saddle-point expansion for the
free energy (\ref{freenDSdirect}) of $U(N)$ gauge theory on $\torus^2$
in the double-scaling limit are given by
\bea
\xi_{2h-1}&=&(4h-3)!\,\sum_{m=1}^h\frac{(-1)^{m-1}\,2^{m+2h-1}}
{(4h-2-m)!}\nonumber\\&&\times\,
\sum_{\stackrel{\scriptstyle\mbf\nu\in\nat_0^h}
{\scriptstyle\sum_kk\,\nu_k=h\,,\,\sum_k\nu_k=m}}~
\prod_{k=1}^h\,\frac1{\nu_k!}\,\left(\frac{\left(2^{2k-1}-1\right)\,
(4k-3)!!~B_{2k}}{(2k)!}\right)^{\nu_k} \ .
\nonumber\eea
\label{xifinal}\end{theorem}
This formula coincides {\it precisely} with Theorem~7.1
of~\cite{eskin1}, whereby the asymptotics of simple Hurwitz numbers
are evaluated directly by involved combinatorial techniques. Here
we have shown that the saddle-point equation provides a very
efficient and much simpler method for extracting these numbers.

\bigskip

\sect{5. Principal moduli spaces of holomorphic differentials} As we will
now demonstrate, Proposition~\ref{Hurwitzprop1} implies that the
double scaling limit of the $U(N)$ gauge theory on $\torus^2$ is
intimately related to the geometry of some very special moduli
spaces~\cite{eskin1,kont2}, well-known in ergodic theory, whose points
are ``counted'' by the strong coupling expansion coefficients of the
previous section. Let $\mathcal{M}_h$ be the moduli space of
(topological classes of) pairs $(\Sigma,\dd\mbf u)$, where $\Sigma$ is
a compact Riemann surface of genus $h$ and $\dd\mbf u$ is a
holomorphic one-form on $\Sigma$ with exactly $2h-2$ simple zeroes
$\{u_i\}_{i=1}^{2h-2}\subset\Sigma$. We call $\mathcal{M}_h$ a
principal moduli space of holomorphic differentials. It can be
coordinatized as follows. Consider the relative homology group
$H_1(\Sigma,\{u_i\};\zed)\cong\zed^{4h-3}$, and choose a basis of
relative one-cycles $\{\gamma_i\}_{i=1}^{4h-3}$ such that $\gamma_i$
for $i=1,\dots,2h$ form a canonical symplectic basis of one-cycles for
the ordinary homology group $H_1(\Sigma;\zed)$, while the open
contours $\gamma_{2h+i}$ for $i=1,\dots,2h-3$ connect the zeroes
$u_{i+1}$ to $u_1$ on $\Sigma$. We define the corresponding period map
$\phi:\mathcal{M}_h\to\complex^{4h-3}$ by $\phi(\Sigma,\dd\mbf
u):=(\,\int_{\gamma_1}\dd\mbf
u\,,\,\dots\,,\,\int_{\gamma_{4h-3}}\dd\mbf u)$. This map defines a
local system of complex coordinates on $\mathcal{M}_h$ which makes it
a complex orbifold of dimension $\dim\mathcal{M}_h=4h-3$. By using the
period map we can also define a smooth measure on the moduli
space $\mathcal{M}_h$ using the pull-back of the Lebesgue measure
$\dd\nu$ from $\complex^{4h-3}$ to $\mathcal{M}_h$ under
$\phi$. However, the total volume of $\mathcal{M}_h$ with respect to
this measure is infinite. To cure this, we restrict to the subspace
$\mathcal{M}_h'\subset\mathcal{M}^{~}_h$ consisting of pairs
$(\Sigma,\dd\mbf u)$ such that the area of the surface $\Sigma$ is~$1$
with respect to the metric defined by the holomorphic one-form
$\dd\mbf u$. Let ${\rm
  C}_N\phi\left(\mathcal{M}_h'\right):=\{t\,\phi(\mathcal{M}_h')~
|~0\leq t\leq\sqrt N\,\}$ be a sequence of cones over
$\phi(\mathcal{M}_h')$ with vertex at the origin of
$\complex^{4h-3}$. We may then define the finite volume of the moduli
space $\mathcal{M}_h'$ by the formula
\beq
{\rm vol}\left(\mathcal{M}_h'\right):=\mbox{$\int_{{\rm
  C}_1\phi(\mathcal{M}_h')}\dd\nu~1$} \ .
\label{volumedef}\eeq

Let us describe the relationship between these moduli spaces and
the double scaling limit of two-dimensional Yang-Mills theory on the
torus $\torus^2$. Consider a branched covering
$\varpi:\Sigma\to\torus^2$ of the torus by a Riemann surface of
genus~$h$, with simple ramification over distinct points
$\{z_i\}_{i=1}^{2h-2}\subset\torus^2$. With $\dd z$ denoting the canonical
holomorphic differential on $\torus^2$, one can use the pull-back
under the covering map to associate the point
$(\Sigma\,,\,\varpi^*(\dd z))\in\mathcal{M}_h$ with simple zeroes
$u_i=\varpi^{-1}(z_i)$ corresponding to the ramification points of the
cover. Conversely, given $(\Sigma,\dd\mbf u)\in\mathcal{M}_h$ we can
define a covering map $\varpi:\Sigma\to\torus^2$ by
$z=\varpi(u):=\int^u\dd\mbf u~~{\rm mod}~\zed^2$. The critical points
of $\varpi$ are precisely the simple zeroes $u_i$ of the holomorphic
differential $\dd\mbf u=\varpi^*(\dd z)$, and $\varpi$ thereby has
simple ramification at $u_i\in\Sigma$. The degree of this covering map
is the area of $\Sigma$ with respect to the metric defined by $\dd\mbf
u$. We have thereby arrived at a one-to-one
correspondence between simple branched covers of the torus, and hence
terms in the chiral Gross-Taylor string expansion, and points in the
principal moduli spaces of holomorphic differentials.

We will now show that the numbers (\ref{volumedef}) are computed by
the strong-coupling saddle-point expansion of the gauge theory
that we obtained in the previous section. The basic idea is
that the counting of points $(\Sigma\,,\,\varpi^*(\dd
z))\in\mathcal{M}_h$ is like counting lattice points $\zed^{2(4h-3)}$
inside subsets of $\real^{2(4h-3)}\cong\complex^{4h-3}$. Using the
definition (\ref{volumedef}) we may compute the volumes of the
principal moduli spaces as
\beq
{\rm vol}\left(\mathcal{M}_h'\right)=\lim_{N\to\infty}\,
\mbox{$\frac1{N^{4h-3}}$}\,\left|\,{\rm C}_N\phi\left(\mathcal{M}_h'\right)
{}~\cap~\left(\zed^{2(4h-3)}+\mbf b\right)\,\right| \ ,
\label{volumelimit}\eeq
where the vector $\mbf b=(b_i)\in\complex^{4h-3}$ has components
$b_i\in\zed^2$ for $i=1,\dots,2h$ while $b_i\neq b_j~~{\rm
  mod}~\zed^2$ for $i,j>2h$ with $i\neq j$. On the other hand, from
the above correspondence it follows that each point of the intersection ${\rm
  C}_N\phi(\mathcal{M}_h')\cap(\zed^{2(4h-3)}+\mbf b)$ corresponds to a simple
branched cover $\varpi$ of $\torus^2$ with winding number $\leq N$, and
thus the volume (\ref{volumelimit}) may be computed via the
asymptotics of simple Hurwitz numbers as
\beq
{\rm vol}\left(\mathcal{M}_h'\right)=\lim_{N\to\infty}\,
\frac1{N^{4h-3}}\,\sum_{n=1}^N\omega_h^n \ .
\label{volumeHurwitz}\eeq
Comparing with (\ref{betahasympts}) and Proposition~\ref{Hurwitzprop1}
we thereby find that the volumes are completely determined in terms of
the coefficients $\xi_{2h-1}$ of the saddle-point solution to be
\beq
{\rm
  vol}\left(\mathcal{M}_h'\right)=\frac{\xi_{2h-1}}{(4h-3)!~(4h-3)}~
\pi^{2h} \ .
\label{volumesaddle}\eeq
Using the formula (\ref{volumesaddle}) and the explicit expansion
(\ref{largeNfreenUNdouble}), the first few volumes can be readily computed and
are summarized in Table~\ref{Hurwitztable}. The general combinatorial
solution is provided by Theorem~\ref{xifinal}. In particular, the
saddle-point computation explicitly demonstrates the rationality
property $\pi^{-2h}~{\rm
  vol}(\mathcal{M}_h')\in\rat$~\cite{eskin1,kont2} and gives a precise
geometrical meaning to the rational numbers that we first encountered
in the weak-coupling expansions of Section~2.
\begin{proposition}
The asymptotic expansion as $\mu\to\infty$ for the free energy of
$U(N)$ gauge theory on $\torus^2$ in the double-scaling limit is the
generating function for the volumes of the principal moduli spaces of
holomorphic differentials given by
$$
\hat{\mathcal{F}}(\mu,N)=N\,\sum_{h=1}^\infty\,
\frac{(4h-3)!}{(\pi\,\mu)^{2h-1}}~{\rm vol}\left(\mathcal{M}_h'
\right)+O\left(\e^{-\mu}\right) \ .
$$
\label{Hurwitzprop2}\end{proposition}

\begin{table}
\begin{center}
\begin{tabular}{|c|c|}\hline \ $h$ \ & \
 {${\rm vol}(\mathcal{M}_h')/\pi^{2h}$} \ \\ \hline\hline
$1$ & {\mbox{$\frac13$}}
\\ \hline $2$ & {\mbox{$\frac2{675}$}} \\ \hline $3$ & {
\mbox{$\frac1{65610}$}} \\ \hline
$4$ & {
\mbox{$\frac{29}{757795500}$}} \\ \hline
$5$ & {
\mbox{$\frac{23357}{422031469860000}$}} \\ \hline
$6$ & {
\mbox{$\frac{16493303}{318258151736124600000}$}}
\\ \hline
\end{tabular}
\end{center}
\caption{\baselineskip=12pt {\it The normalized volumes of the
    principal moduli spaces of holomorphic differentials up to genus
    $h=6$.}}
\label{Hurwitztable}\end{table}

\bigskip

\sect{6. String theory interpretation} Let us conclude by briefly
discussing some of the implications that our
results have on the string expansion of two-dimensional Yang-Mills
theory. As we have seen, only the chiral part of the Gross-Taylor
series contributes in the double scaling limit of $U(N)$
gauge theory on $\torus^2$, and so the effective string theory is
necessarily chiral. Alternatively, we may regard this fact as being
the statement that the holomorphic and anti-holomorphic sectors are
identified with one another, so that the expansion is in terms of {\it
  open} strings. Being a chiral series, the original strong-coupling
expansion is given as a sum over {\it all} Young diagrams, labelling
representations of the infinite unitary group $U(\infty)$, which is
the gauge group of a {\it noncommutative} gauge theory. In the strong
coupling regime, the double scaling limit produces the asymptotic
Hurwitz numbers, of branched covering maps to $\torus^2$ in the limit
of infinite winding number, again reflecting the open nature of the
string degrees of freedom. This latter feature also implies that the
toroidal spacetime is effectively decompactified onto the plane in the
double scaling limit, corresponding to the equi-anharmonic limit
$\tau\to\ii\infty$ of the underlying elliptic curve. All of these
features suggest that the proper setting for understanding the physics
of the double scaling limit is through the fluxon expansion of gauge
theory on the noncommutative plane. This is indeed the case and will
be analysed in more detail in~\cite{inprep}.

Let us now note that the Gross-Taylor series is an expansion in
$\frac1N$, i.e. it is {\it perturbative} in the string coupling $g_s$,
and as such it ignores non-perturbative corrections of the form
$\e^{-N\,\lambda}$. These contributions have a natural
interpretation~\cite{LMR1} as coming from (Euclidean) D1-branes of
tension $T_1=\frac1{\pi\,\alpha'g_s}$ which wrap around the target
space torus $\torus^2$ without foldings. The new double scaling limit
of this paper captures the D-string contributions which are of
order $\e^{-\mu}$, with the double scaled coupling constant related to
the brane tension. These contributions play an important
role in ensuring regularity of the complete free energy given by
(\ref{bella}), as their collective behaviour must eliminate the
divergences which are just artifacts of the strong-coupling
approximation. We can now rewrite the strong-coupling limit of the
free energy (\ref{freenDSdirect}) in terms of open string parameters
as
\beq
\mathcal{Z}_{\rm str}(g_s,T_1A):=\lim_{\mu\to\infty}\,\hat
{\mathcal F}(\mu,N)=\frac1{g_s}\,\sum_{k=1}^\infty2^{2k-1}\,
(4k-3)!~\frac{{\rm vol}\left(\mathcal{M}_k'\right)}{(T_1A)^{2k-1}} \ .
\label{openstrdouble}\eeq
We interpret this expansion as {}{a remnant of the resummation of the D1-branes
provided by (\ref{bella})}. The
effective action (\ref{openstrdouble}) is of order $1/g_s$, thereby
representing an open string {\it disk} amplitude, and so the
double scaling limit can be interpreted as a theory of
D1-branes at tree-level in open string perturbation theory. The
truncation of the dynamics to tree level is presumably related to the
fact that the strong coupling regime of the double scaling gauge
theory is described by some sort of topological open string
theory. The natural appearence of moduli space volumes strongly suggests
that an explicit realization of the string partition function
generically as an Euler sigma-model should be possible, so that the
string path integral localizes in the usual way onto finite dimensional
moduli spaces~\cite{CMR1}. The double scaling limit replaces the
counting of holomorphic maps $\Sigma\to\torus^2$, arising in the
't~Hooft limit of QCD$_2$ on the torus, by the counting of holomorphic
differentials on complex curves $\Sigma$. The action principle for
this open string theory is given by the topologically twisted
$\mathcal{N}=2$ superconformal field theory coupled to gravity that
describes the closed string expansion of two-dimensional Yang-Mills
theory~\cite{bcov}, by taking the torus target space to be
equi-anharmonic after a modular transformation. More details will be
given in~\cite{inprep}.

\bigskip

\noindent
{\bf Acknowledgments}

\noindent
We thank A.~Polychronakos for valuable comments on the manuscript, and
G.~Cicuta, F.~Colomo, Ph.~Flajolet, S.~Ramgoolam and
J.~Wheater for helpful discussions. The work of R.J.S. was supported
in part by an Advanced Fellowship from the Particle Physics and
Astronomy Research Council~(U.K.).

\end{document}